# Tomonaga-Luttinger Liquid Exponents of Correlation Functions around Fermi points in the Hubbard Model

Nelson O. Nenuwe[1*]    John O.A. Idiodi[2]
1.Department of Physics, Federal University of Petroleum Resources, P.M.B 1221, Effurun, Delta State, Nigeria
2.Department of Physics, University of Benin, P.M.B. 1154, Benin City, Edo State, Nigeria
* E-mail of the corresponding author: favorednelso@gmail.com

**Abstract**
The correlation functions of one-dimensional Hubbard model in the presence of external magnetic field was investigated through the conformal field technique. The long distance behaviour of the correlation functions and their critical exponents for the model in the presence of a magnetic field are established by solving the dressed charge matrix equations and setting the number of occupancies $N_{c,s}^{\pm}$ to one, as an alternative to the usual zero often used by authors in literatures. Our result shows the critical exponents of the correlation function grows monotonically with magnetic field and reduces to definite values at zero magnetic field around various Fermi points.

## 1. Introduction

Over two decades ago Frahm and Korepin (1990) introduced the calculation of critical exponents for the one-dimensional (1D) Hubbard model, using the finite size scaling and the principle of conformal field theory (CFT). This enabled theorists to explore the physics of 1D correlated electron systems. Notwithstanding significant works, the understanding of the behaviour of correlated electron systems is not yet complete. In one dimension the Hubbard Hamiltonian provides opportunity to study correlation effects in 1D models (Lieb and Wu, 1968; Woynarovich, 1989) and the correlation functions decay as power of the distance (Nenuwe and Akpojotor, 2015). It is the calculation of the critical exponents characterizing this power-law behaviour that have attracted constant theoretical interest. Outstanding results in this field have been obtained from conformal field techniques, perturbation calculations and renormalization group methods in different models (Frahm and Korepin, 1990; Parola and Sorella, 1990; Finkel'shtein, 1977; Luther and Peschel, 1975). The progress made in the understanding of critical phenomena in quantum systems as a result of conformal invariance (Belavin *et al*., 1984) have provided great insights to the problem of calculation of these critical exponents. Although, interacting 1D quantum systems might carry countless low-energy excitations, with linear dispersion relations, but with different Fermi velocities, so the systems are not Lorentz invariant (Kawakami and Yang, 1991). When the motion of these excitations are decoupled, one can now apply the CFT (Izergin *et al*., 1989). Usually, in the application of the conformal field techniques, the non-negative integer $N_{c,s}^{\pm}$ characterizing particle-hole excitations is always taken as zero, but in this paper we shall calculate the electron field correlation function and the density-density correlation function around the Fermi points $k_F$, $2k_F$, $3k_F$, $4k_F$, $5k_F$, $6k_F$, $7k_F$, $8k_F$, $9k_F$, $11k_F$ and $13k_F$ by setting the parameter $N_{c,s}^{\pm}$ to one, and investigate how this affects the conformal dimensions and critical exponents of the correlation functions. This paper is organized as follows. In section 2 we review the Bethe Ansatz equations of the Hubbard model and the analytic form of the correlation functions predicted by CFT is given. The dressed charge matrix elements are also calculated with the Wiener-Hopf technique and these elements are used to obtain the magnetic field dependence of the conformal dimensions. The long-distance behaviour of the electron field and density-density correlation functions and their unusual exponents for small magnetic field are calculated in section 3. The electron field correlation function in momentum space and their Tomonaga-Luttinger (TL) liquid behaviour is examined in section 4. Finally, section 5 is devoted to discussion and conclusion.

## 2. The Dressed Charge Matrix and the Hubbard Model

The Hubbard model is basically the simplest model describing interacting spin-1/2 fermions in many-body physics. In the presence of magnetic field it is defined by the Hamiltonian (Penc and Solyom, 1993)

$$H = -\sum_{j,\sigma}\left(c_{j+1,\sigma}^{\dagger}c_{j,\sigma} + c_{j,\sigma}^{\dagger}c_{j+1,\sigma}\right) + u\sum_{j}n_{j,\uparrow}n_{j,\downarrow} - \mu\sum_{j}\left(n_{j,\uparrow}+n_{j,\downarrow}\right) - \frac{H}{2}\sum_{j}\left(n_{j,\uparrow}-n_{j,\downarrow}\right), \quad (1)$$

where $c_{j,\sigma}^{\dagger}\left(c_{j,\sigma}\right)$ is the creation (annihilation) operator with electron spin σ at site *j* and $n_{j,\sigma}=c_{j,\sigma}^{\dagger}c_{j,\sigma}$ is the number operator. *u* is the on-site Coulomb repulsion, *μ* is the chemical potential and *H* is the external magnetic field. The hopping integral *t*=1. Lieb and Wu (1968) has solved Eqn. (1) exactly and obtained the Bethe Ansatz equations

$$N k_j = 2\pi I_j + \sum_{\beta=1}^{N_s} 2\tan^{-1}\left(\frac{\sin k_j - \lambda_\beta}{u}\right) \quad (2)$$

$$\sum_{j=1}^{N_c} 2\tan^{-1}\left(\frac{\lambda_\alpha - \sin k}{u}\right) = 2\pi J_\alpha + \sum_{\beta=1}^{N_s} 2\tan^{-1}\left(\frac{\lambda_\alpha - \lambda_\beta}{u}\right) \quad (3)$$

Where the quantum number $I_j$ and $J_\alpha$ are integers or half-odd integer, $N_c = N_\uparrow + N_\downarrow$ with $N_\uparrow$ and $N_\downarrow$ being the number of electrons with spin up and down, and $N_s = N_\downarrow$ down spins are characterized by the momenta $k_j$ of holons and rapidities $\lambda_\alpha$ of spinons.

In the thermodynamic limit, with continuous momentum and rapidity variables, the Lieb-Wu equations become integral equations for the ground state distribution functions of momenta $\rho_c(k)$ and of rapidities $\rho_s(\lambda)$, obeying the equations

$$\rho_c(k) = \frac{1}{2\pi} + \frac{\cos k}{2\pi}\int_{-\lambda_0}^{\lambda_0} a_1(\sin k - \lambda)\rho_s(\lambda)d\lambda,$$
$$\rho_s(\lambda) = \frac{1}{2\pi}\int_{-k_0}^{k_0} a_1(\lambda - \sin k)\rho_c(k)dk - \frac{1}{2\pi}\int_{-\lambda_0}^{\lambda_0} a_2(\lambda - \mu)\rho_s(\mu)d\mu \quad (4)$$

The state corresponding to the solution of Eqns. (2) and (3) has energy and momentum given by

$$E_n(I,D) - E_0 = \frac{2\pi}{N}v_c(\Delta_c^+ + \Delta_c^-) + \frac{2\pi}{N}v_s(\Delta_s^+ + \Delta_s^-) + O(N^{-1}) \quad (5)$$

$$P(I,D) - P_0 = (2\pi - 2k_{F,\uparrow} - k_{F,\downarrow})D_c + (2\pi - 2k_{F,\uparrow})D_s + \frac{2\pi}{N}(\Delta_c^+ - \Delta_c^- + \Delta_s^+ - \Delta_s^-) \quad (6)$$

Where the conformal dimensions are given by

$$2\Delta_c^\pm(\Delta N, D) = \left(Z_{cc}D_c + Z_{sc}D_s \pm \frac{Z_{ss}\Delta N_c - Z_{cs}\Delta N_s}{2\det Z}\right)^2 + 2N_c^\pm \quad (7)$$

$$2\Delta_s^\pm(\Delta N, D) = \left(Z_{cs}D_c + Z_{ss}D_s \pm \frac{Z_{cc}\Delta N_s - Z_{sc}\Delta N_c}{2\det Z}\right)^2 + 2N_s^\pm \quad (8)$$

The positive integers $N_{c,s}^\pm$, for holon and spinon describes particle-hole excitations, with $N_{c,s}^+(N_{c,s}^-)$ being the number of occupancies that a particle at the right (left) Fermi level jumps to, $\Delta N_c(\Delta N_s)$ represents the change in the number of electrons (down-spin) with respect to the ground state, $D_c$ represents the number of particles which transfer from one Fermi level of the holon to the other and $D_s$ represents the number of particles which transfer from one Fermi level of the spinon to the other, and both $D_c$ and $D_s$ are either integer or half-odd integer values. Finally, the dressed charge matrix Z describing anomalous behaviour of critical exponents is given by

$$Z = \begin{pmatrix} Z_{cc} & Z_{cs} \\ Z_{sc} & Z_{ss} \end{pmatrix}, \quad (9)$$

and the elements are defined by the solutions of the following coupled integral equations

$$Z_{cc}(k) = 1 + \int_{-\lambda_0}^{\lambda_0} d\lambda\, a_1(k-\lambda)\xi_{cs}(\lambda) \quad (10)$$

$$Z_{cs}(\lambda) = \int_{-k_0}^{k_0} dk\, a_1(\lambda - k)Z_{cc}(k) - \int_{-\lambda_0}^{\lambda_0} d\mu\, a_2(\lambda - \mu)Z_{cs}(\mu) \quad (11)$$

$$Z_{sc}(k) = \int_{-\lambda_0}^{\lambda_0} d\lambda\, a_1(k-\lambda)Z_{ss}(\lambda) \quad (12)$$

$$Z_{ss}(\lambda) = 1 + \int_{-k_0}^{k_0} dk\, a_1(\lambda - k)Z_{sc}(k) - \int_{-\lambda_0}^{\lambda_0} d\mu\, a_2(\lambda - \lambda')Z_{cs}(\mu) \quad (13)$$

Where the kernel is defined as

$$a_n(x) = \frac{2}{\pi}\frac{n\,u}{(n\,u)^2 + x^2} \quad (14)$$

The values of $\lambda_0$ and $k_0$ are usually fixed by

$$n_c = \int_{-k_0}^{k_0}\rho_c(k)dk = \int_{-k_0}^{k_0}\frac{Z_{cc}(k)}{2\pi}dk \quad (15)$$

$$n_s = \int_{-\lambda_0}^{\lambda_0}\rho_s(\lambda)d\lambda = \int_{-k_0}^{k_0}\frac{Z_{cs}(\lambda)}{2\pi}d\lambda = \int_{-k_0}^{k_0}\frac{Z_{sc}(k)}{2\pi}dk \quad (16)$$

For small magnetic field we solve the dressed charge matrix equations by Wiener-Hopf technique (Fabian et al., 2005; Yang and Yang, 1966) for terms up to order $1/u$ in the strong coupling limit. With Eqn. (16), we write (13) as

$$Z_{ss}(\lambda) = 1 + a_1(\lambda)\int_{-k_0}^{k_0} Z_{sc}(k)\,dk - \int_{-\lambda_0}^{\lambda_0} a_2(\lambda-\mu)\,Z_{ss}(\mu)d\mu$$
$$= 1 + 2\pi n_s a_1(\lambda) - \int_{-\lambda_0}^{\lambda_0} a_2(\lambda-\mu)\,Z_{ss}(\mu)d\mu \qquad (17)$$

Fourier transforming (17), we obtain

$$Z_{ss}(\lambda) = \tfrac{1}{2} + 2\pi n_s a_1(\lambda) - \int_{|\mu|\geq\lambda_0} K(\lambda-\mu)Z_{ss}(\mu)d\mu \qquad (18)$$

Where the kernels are given by

$$s(\lambda) = \frac{1}{2u\cosh(\pi\lambda/u)}$$
$$K(\lambda) = \frac{1}{2}\int_{-\infty}^{\infty}\frac{\exp(-i\omega\lambda)}{1+\exp(\omega u)}d\omega \qquad (19)$$

We solve Eqn. (18) by introducing the function

$$y(\lambda) = Z_{ss}(\lambda+\lambda_0), \qquad (20)$$

and expanding it as

$$y(\lambda) = \sum_{n=o}^{\infty} y_n(\lambda) \qquad (21)$$

Where $y_n(\lambda)$ are defined as the solutions of the Wiener-Hopf equations

$$y_n(\lambda) = g_n(\lambda) + \int_0^{\infty} K(\lambda-\mu)y_n(\mu)d\mu \qquad (22)$$

$$g_n(\lambda) = \int_0^{\infty} K(\lambda+\mu+2\lambda_0)y_{n-1}(\mu)d\mu, \quad n\geq 1 \qquad (23)$$
$$g_0(\lambda) = \tfrac{1}{2} + 2\pi n_s s(\lambda+\lambda_0)$$

The driving terms $g_n(\lambda)$ and the solutions $y_n(\lambda)$ becomes smaller as $n$ increases because $\lambda$ is large. Our procedure follows Fabian et al. (2005). Assuming the function $y_{n-1}(\lambda)$ and $g_n(\lambda)$ are known. We define

$$\tilde{y}_n^+(\omega) = \int_0^{\infty}\exp(i\omega\lambda)y_n(\lambda)d\lambda$$
$$\tilde{y}_n^-(\omega) = \int_{-\infty}^{0}\exp(i\omega\lambda)y_n(\lambda)d\lambda \qquad (24)$$

Where the functions $\tilde{y}_n^{\pm}(\omega)$ are analytic on the upper and lower planes respectively, with

$$\tilde{y}(\omega) = \tilde{y}_n^+(\omega) + \tilde{y}_n^-(\omega) \qquad (25)$$

Also we assume

$$\tilde{y}_n^{\pm}(\infty) = \tilde{g}(\infty) = 0 \qquad (26)$$

In terms of these functions we express the Fourier transform of Eqn. (23) as

$$\tilde{g}_n(\omega) = \frac{\tilde{y}_n^+(\omega)}{1+\exp(-2u|\omega|)} + \tilde{y}_n^-(\omega) \qquad (27)$$

Where $\tilde{g}_n(\omega)$ is the Fourier transform of $g_n(\lambda)$. Now we split Eqn. (23) into the sum of two parts that are analytical and non-zero in the upper and lower half planes. To obtain this we use the factorization

$$1+\exp(-2u|\omega|) = G^+(\omega)G^-(\omega) \qquad (28)$$

$$G^+(\omega) = G^-(-\omega)\frac{\sqrt{2\pi}}{\Gamma(\tfrac{1}{2}-\tfrac{iu\omega}{\pi})}\left(-\frac{iu\omega}{\pi}\right)^{-\tfrac{iu\omega}{\pi}}e^{\tfrac{iu\omega}{\pi}} \qquad (29)$$

With $G^{\pm}(\omega)$ being analytic and non-zero in the upper and lower half planes respectively and are normalized as

$$\lim_{\omega\to\infty} G^{\pm}(\omega) = 1 \qquad (30)$$

Useful special function of $G^{\pm}(\omega)$ are

$$G^{\pm}(0) = \sqrt{2}$$
$$G^{\pm}\left(\frac{i\pi}{2u}\right) = G^{\pm}\left(-\frac{i\omega}{2u}\right) = \sqrt{\frac{\pi}{e}} \qquad (31)$$

Using Eqns. (27) and (28), we obtain

$$\frac{\tilde{y}_n^+(\omega)}{G^+(\omega)} + G^-(\omega)\tilde{y}_n^-(\omega) = G^-(\omega)\tilde{g}_n(\omega) \qquad (32)$$

Decompose the right hand side of Eqn. (32) into the sum of two functions

$$G^-(\omega)\tilde{g}_n(\omega) = Q_n^+(\omega) + Q_n^-(\omega) \qquad (33)$$

This implies that

$$\tilde{y}_n^+(\omega) = G^+(\omega)Q_n^+(\omega)$$
$$\tilde{y}_n^-(\omega) = \frac{Q_n^-(\omega)}{G^-(\omega)} \qquad (34)$$

To obtain the solution of Eqn. (22) for $y_0(\lambda)$, we set the driving term to be

$$\tilde{g}(\omega) = 2\pi\delta(\omega) + \frac{b\, e^{-i\lambda_0\omega}}{2\cosh(u\omega)} \qquad (35)$$

We decompose the first term by using

$$2\pi\delta(\omega) = i\left(\frac{1}{\omega+i\varepsilon} - \frac{1}{\omega-i\varepsilon}\right), \qquad (\varepsilon \to 0) \qquad (36)$$

The second term of Eqn. (35) is meromorphic function of $\omega$ with simple poles located at

$$\omega_n = \frac{i\pi}{2u}(2n+1)$$
$$\omega_0 = \frac{i\pi}{2u}, \quad \omega_1 = \frac{3i\pi}{2u}, \quad \omega_2 = \frac{5i\pi}{2u}, \ldots \qquad (37)$$

Note, there is no pole at $\omega = 0$. The decomposition of the factor $1/\cosh(u\omega)$ gives

$$\frac{1}{\cosh(u\omega)} = A^+(\omega) + A^-(\omega) \qquad (38)$$

$$A^+(\omega) = \frac{i}{u}\sum_{n=0}^{\infty}\frac{(-1)^n}{\omega+\omega_n}$$
$$A^-(\omega) = \frac{1}{\cosh(u\omega)} - \frac{i}{u}\sum_{n=0}^{\infty}\frac{(-1)^n}{\omega+\omega_n} \qquad (39)$$

Using Eqn. (39) we can express the function $f^-(\omega)/\cosh(u\omega)$, for any function $f^-(x)$ that is analytic and bounded in the lower half-plane as the sum of two functions $\chi^\pm(\omega)$ analytic in the upper/lower half-plane

$$\frac{f^-(\omega)}{\cosh(u\omega)} = \chi^+(\omega) + \chi^-(\omega) \qquad (40)$$

$$\chi^+(\omega) = \frac{i}{u}\sum_{n=0}^{\infty}\frac{(-1)^n f^-(-\omega_n)}{\omega+\omega_n}$$
$$\chi^-(\omega) = \frac{f^-(-\omega_n)}{\cosh(u\omega)} - \frac{i}{u}\sum_{n=0}^{\infty}\frac{(-1)^n f^-(-\omega_n)}{\omega+\omega_n} \qquad (41)$$

Applying the formula Eqn.(41) to Eqns. (35) and (33), we obtain

$$\tilde{g}(\omega) = ai\left(\frac{1}{\omega+i\varepsilon} - \frac{1}{\omega-i\varepsilon}\right) + \frac{b}{2}[\chi^+(\omega) + \chi^-(\omega)] \qquad (42)$$

$$\tilde{g}(\omega) = \frac{ai}{\omega+i\varepsilon} + \frac{bi}{2u}\sum_{n=0}^{\infty}\frac{(-1)^n e^{-i\lambda_0\omega}}{\omega+\omega_n} - \frac{ai}{\omega-i\varepsilon} + \frac{b}{2}\frac{e^{-i\lambda_0\omega}}{\cosh(u\omega)} - \frac{bi}{2u}\sum_{n=0}^{\infty}\frac{(-1)^n e^{-i\lambda_0\omega}}{\omega+\omega_n} \qquad (43)$$

Now,

$$G^-(\omega)\tilde{g}(\omega) = \frac{aiG^-(\omega)}{\omega+i\varepsilon} + \frac{bi}{2u}\sum_{n=0}^{\infty}\frac{(-1)^n G^-(\omega)e^{-i\lambda_0\omega}}{\omega+\omega_n} - \frac{aiG^-(\omega)}{\omega-i\varepsilon} + \frac{b}{2}\frac{G^-(\omega)e^{-i\lambda_0\omega}}{\cosh(u\omega)} - \frac{bi}{2u}\sum_{n=0}^{\infty}\frac{(-1)^n G^-(\omega)e^{-i\lambda_0\omega}}{\omega+\omega_n} \qquad (44)$$

$$\equiv Q_n^+(\omega) + Q_n^-(\omega)$$

Therefore,

$$Q_n^-(\omega) = -\frac{aiG^-(-\omega_n)}{\omega-i\varepsilon} + \frac{b}{2}\frac{G^-(-\omega_n)e^{-i\lambda_0\omega}}{\cosh(u\omega)} - \frac{bi}{2u}\sum_{n=0}^{\infty}\frac{(-1)^n G^-(-\omega_n)e^{-i\lambda_0\omega_n}}{\omega+\omega_n} \qquad (45)$$

$$Q_n^+(\omega) = \frac{aiG^-(-\omega_n)}{\omega+i\varepsilon} + \frac{bi}{2u}\sum_{n=0}^{\infty}\frac{(-1)^n G^-(-\omega_n)e^{-i\lambda_0\omega}}{\omega+\omega_n} \qquad (46)$$

For $n = 0$

$$Q_0^+(\omega) = \frac{aiG^-(0)}{\omega + i\varepsilon} + \frac{bi}{2u} \frac{G^-\left(\frac{-\pi i}{2u}\right) e^{\frac{\lambda_0 \pi}{2u}}}{\omega + \frac{\pi i}{2u}} \qquad (47)$$

The functions $\tilde{y}_0^\pm(\omega)$ are obtained by using Eqn. (34)

$$\tilde{y}_0^+(\omega) = G^+(\omega) \left( \frac{aiG^-(0)}{\omega + i\varepsilon} + \frac{bi}{2u} \frac{G^-\left(\frac{-\pi i}{2u}\right) e^{\frac{\lambda_0 \pi}{2u}}}{\omega + \frac{\pi i}{2u}} \right) \qquad (48)$$

From Eqn. (23) for $n = 0$, by setting $a = 1/2$, $b = 2\pi n_s$ in Eqn. (48), we obtain

$$\tilde{y}_0^+(\omega) = G^+(\omega) \left( \frac{1}{2} \frac{iG^-(0)}{\omega + i\varepsilon} + \frac{\pi n_s i}{u} \frac{G^-\left(\frac{-\pi i}{2u}\right) e^{\frac{\lambda_0 \pi}{2u}}}{\omega + \frac{\pi i}{2u}} \right) + O(H^2) \quad (49)$$

By definition

$$y(0) = \frac{1}{2\pi} \int_{-\infty}^{\infty} \tilde{y}^+(\omega) e^{-i\omega\varepsilon} d\omega = -i \lim_{\omega \to \infty} \omega \tilde{y}^+(\omega) \qquad (50)$$

$$\lambda_0 \approx \frac{2u}{\pi} \ln\left(\frac{H_0}{H}\right)$$

$$H_0 = \sqrt{\frac{\pi^3}{2e}} H_c \qquad (51)$$

$$H_c = \frac{4\pi^2 n_c^3}{3u}\left(1 - \frac{\pi^2 n_c^2}{5u^2}\right); \quad u \gg 1$$

Where $H$ is magnetic field, $H_c$ is critical field, $u$ strong coupling, $H_0$ magnetic field at zero temperature and $\lambda_0$ corresponds to Fermi points. Combining the result Eqn. (49) with Eqn. (50), we obtain the first order contribution to $Z_{ss}$ as follows

$$-i \lim_{\omega \to \infty} \omega \tilde{y}^+(\omega) = -i \lim_{\omega \to \infty} \omega G^+(\omega) \left( \frac{1}{2} \frac{iG^-(0)}{\omega + i\varepsilon} + \frac{\pi n_s i}{u} \frac{G^-\left(\frac{-\pi i}{2u}\right) e^{\frac{\lambda_0 \pi}{2u}}}{\omega + \frac{\pi i}{2u}} \right) \quad (52)$$

As $\varepsilon \to 0$, we use Eqns. (30) and (31) on Eqn. (52) to obtain

$$y_0(0) = -i\omega \left( \frac{1}{2} \frac{i\sqrt{2}}{\omega} + \frac{\pi n_s i}{u} \frac{\sqrt{\frac{\pi}{e}} e^{\frac{\lambda_0 \pi}{2u}}}{\omega + \frac{\pi i}{2u}} \right) \qquad (53)$$

Simplifying further, we obtain

$$y_0(0) = \frac{\sqrt{2}}{2} + \frac{\pi n_s}{u} \frac{\sqrt{\frac{\pi}{e}}}{1 + \frac{\pi i}{2u\omega}} \exp\left(-\frac{\pi}{2u} \cdot \frac{2u}{\pi} \ln(H_0/H)\right)$$

$$= \frac{\sqrt{2}}{2} + \frac{\pi n_s}{u} \frac{\sqrt{\frac{\pi}{e}}}{1 + \frac{\pi i}{\infty}} \frac{H}{H_0}, \quad \text{since } e^{\ln x} = x \quad (54)$$

$$= \frac{\sqrt{2}}{2} + \frac{\pi n_s \sqrt{\frac{\pi}{e}}}{u} \left(\frac{H}{H_0}\right)$$

Using Eqn. (51), we obtain

$$y_0(0) = \frac{\sqrt{2}}{2} + \frac{\pi n_s \sqrt{\frac{\pi}{e}}}{u} \left(\frac{H}{\sqrt{\frac{\pi^3}{2e}} H_c}\right) \qquad (55)$$

$$= \sqrt{2}\left(\frac{1}{2} + \frac{n_s}{u} \frac{H}{H_c}\right) + O(H^2)$$

Next, the second order contribution to $y(0) = Z_{ss}(\lambda_0)$ is obtained by taking the Fourier transform of Eqn. (23) for $n = 1$.

$$g_1(\lambda) = \int_0^\infty K(\lambda + \mu + 2\lambda_0) y_0(\mu) d\mu$$

$$\tilde{g}_1(\omega) = \frac{\exp(-2i\lambda_0 \omega) \tilde{y}_0^+(-\omega)}{1 + \exp(2u|\omega|)} \qquad (56)$$

From Eqn. (28)

$$\frac{1}{1+\exp(2u|\omega|)} = 1 - \frac{1}{G^+(\omega)G^-(\omega)} \qquad (57)$$

$$\tilde{g}_1(\omega) = \exp(-2i\lambda_0\omega)\tilde{y}_0^+(-\omega)\left(1 - \frac{1}{G^+(\omega)G^-(\omega)}\right) \qquad (58)$$

From Eqn. (33),

$$G^-(\omega)\tilde{g}_1(\omega) = Q_1^+(\omega) + Q_1^-(\omega) \qquad (59)$$

We have decomposed $G^-(\omega)\tilde{g}_1(\omega)$ into $Q_1^{\pm}(\omega)$ which is analytic in the upper and lower half-planes. $Q_1^+(\omega)$ is given by

$$Q_1^+(\omega) = \frac{1}{2\pi i}\int_{-\infty}^{\infty}\frac{\exp(-2i\lambda_0 x)\tilde{y}_0^+(-x)}{x-\omega-i\varepsilon}\frac{1}{G^+(x)}dx \qquad (60)$$

Where $\varepsilon$ is a small positive constant. $G^+(x)$ has a branch cut along the negative imaginary axis and by deforming the contour of integration we rewrite Eqn. (58) as

$$Q_1^+(\omega) = \frac{1}{2\pi i}\int_0^{\infty}\frac{\exp(-2i\lambda_0 x)\tilde{y}_0^+(ix)}{x-i\omega}\left(\frac{1}{G^+(-ix-\varepsilon)} - \frac{1}{G^+(-ix+\varepsilon)}\right)dx \qquad (61)$$

From Eqn. (29), as $\omega \to ix$

$$\frac{1}{G^+(-ix-\varepsilon)} = \frac{\Gamma(\frac{1}{2} - \frac{ux}{\pi})}{\sqrt{2\pi}}\left(\frac{ux}{\pi}\right)^{\frac{iux}{\pi}}e^{\frac{iux}{\pi}} \qquad (62)$$

$$Q_1^+(\omega) = \frac{1}{(2\pi)^{\frac{3}{2}}i}\int_0^{\infty}\frac{e^{-2\lambda_0 x}\tilde{y}_0^+(ix)}{x-i\omega}\Gamma(\frac{1}{2}-\frac{ux}{\pi})\left(\frac{ux}{\pi}\right)^{\frac{iux}{\pi}}\left(e^{\frac{iux}{\pi}} - e^{\frac{iux}{\pi}}\right)dx \qquad (63)$$

Since, $\sin x = (e^x - e^{-x})/2i$

$$Q_1^+(\omega) = \frac{2i}{(2\pi)^{\frac{3}{2}}i}\int_0^{\infty}\frac{e^{-2\lambda_0 x}\tilde{y}_0^+(ix)}{x-i\omega}\left(\frac{ux}{\pi}\right)^{\frac{iux}{\pi}}\Gamma(\frac{1}{2}-\frac{ux}{\pi})\sin(ux)dx \qquad (64)$$

For $x > 0$ the integrand rapidly decrease because $\lambda_0 \gg 1$, and hence the integral is approximated by expanding the terms other than $\exp(-2\lambda_0 x)$ around $x=0$. Therefore, we obtain

$$Q_1^+(\omega) \approx \frac{1}{2\pi}\int_0^{\infty}\frac{e^{-2\lambda_0 x}}{-i\omega}\left(\frac{2u}{\sqrt{2}} + O(x)\right)dx$$
$$= \frac{1}{-i\omega}\left(\frac{u}{2\sqrt{2}\pi\lambda_0} + O\left(\frac{1}{\lambda_0^2}\right)\right) \qquad (65)$$

From Eqn. (34), we obtain

$$\tilde{y}_1^+(\omega) = \frac{G^+(\omega)}{-i\omega}\left(\frac{u}{2\sqrt{2}\pi\lambda_0} + O\left(\frac{1}{\lambda_0^2}\right)\right) \qquad (66)$$

Using

$$y_1(0) = -i\lim_{\omega\to\infty}\omega\,\tilde{y}_1^+(\omega) \quad \text{and} \quad \lim_{\omega\to\infty}G^{\pm}(\omega) = 1 \quad , \qquad (67)$$

we obtain

$$y_1(0) = \frac{u}{2\sqrt{2}\pi\lambda_0} + O\left(\frac{1}{\lambda_0^2}\right) \qquad (68)$$

From Eqn. (51)

$$y_1(0) = \frac{1}{4\sqrt{2}}\frac{1}{\ln(H_0/H)} + O\left(\frac{1}{(\ln H_0/H)^2}\right)$$
$$y_1(0) = \frac{\sqrt{2}}{8\ln(H_0/H)} + O\left(\frac{1}{(\ln H_0/H)^2}\right) \qquad (69)$$

Therefore, with Eqns. (55) and (69), we obtain

$$Z_{ss}(\lambda_0) = \sqrt{2}\left(\frac{1}{2} + \frac{n_s}{u}\frac{H}{H_c} + \frac{1}{8\ln(H_0/H)}\right) + O\left(\frac{1}{(\ln H_0/H)^2}\right) \quad (70)$$

Now to evaluate the dressed charge matrix element $Z_{sc}(k_0)$, we take the Fourier transform of eqns. (16) and (18) and obtain

$$Z_{sc}(k) = \tfrac{1}{2} + 2\pi n_s K(k) - \int_{|\mu| \geq \lambda_0} K(k-\lambda) Z_{ss}(\lambda) d\lambda \quad (71)$$

Applying the same process in the determination of Eqn. (70), we obtain

$$Z_{sc}(k_0) = \frac{1}{2} + \frac{n_s \ln 2}{u} - \frac{2}{\pi^2}\frac{H}{H_c} + O\left(\frac{H}{H_c \ln(H_0/H)}\right) \quad (72)$$

Similarly, with the same process, we obtain the other two elements of the dressed charge matrix as

$$Z_{cc}(k_0) = 1 + \frac{n_c \ln 2}{u} - \frac{2n_c}{\pi^2 u}\left(\frac{H}{H_c}\right)^2 + O\left(\frac{H^2}{H_c^2[\ln(H_0/H)]^2}\right) \quad (73)$$

and

$$Z_{cs}(\lambda_0) = \frac{\sqrt{2}n_c}{u}\frac{H}{H_c} + O\left(\frac{H}{H_c[\ln(H_0/H)]^2}\right) \quad (74)$$

From Eqn. (16) together with the property that $Z_{sc}(k) \approx Z_{sc}(k_0) + O(\frac{1}{u^2})$ for $u \gg 1$ and $k_0 \approx \pi n_c/(1+\frac{n_c \ln 2}{u})$, the down-spin density $n_s$ is obtained as

$$n_s = \frac{n_c}{2} - \frac{2n_c}{\pi^2}\frac{H}{H_c} \quad (75)$$

Using Eqn. (75) on Eqns. (70) and (72), we obtain the dressed charge matrix equations as

$$Z_{cc}(k_0) = 1 + \frac{n_c}{u}\left(\ln 2 - \frac{2}{\pi^2}\left(\frac{H}{H_c}\right)^2\right) + O\left(\frac{H^2}{H_c^2[\ln(H_0/H)]^2}\right) \quad (76)$$

$$Z_{cs}(\lambda_0) = \frac{\sqrt{2}n_c}{u}\frac{H}{H_c} + O\left(\frac{H}{H_c[\ln(H_0/H)]^2}\right) \quad (77)$$

$$Z_{sc}(k_0) = \frac{1}{2} - \frac{2}{\pi^2}\frac{H}{H_c} + \frac{n_c \ln 2}{u}\left(\frac{1}{2} - \frac{2}{\pi^2}\frac{H}{H_c}\right) + O\left(\frac{H}{H_c \ln(H_0/H)}\right) \quad (78)$$

$$Z_{ss}(\lambda_0) = \sqrt{2}\left(\frac{1}{2} + \frac{n_c}{u}\left(\frac{1}{2}\frac{H}{H_c} - \frac{2}{\pi^2}\left(\frac{H}{H_c}\right)^2\right) + \frac{1}{8\ln(H_0/H)}\right) + O\left(\frac{1}{(\ln H_0/H)^2}\right) \quad (79)$$

At half-filling $n_c = 1$, and by neglecting corrections to order $(1/u)$, the elements of the dressed charge become

$$Z_{cc}(k_0) = 1 \quad (80)$$

$$Z_{cs}(\lambda_0) = 0 \quad (81)$$

$$Z_{sc}(k_0) = \frac{1}{2} - \frac{2}{\pi^2}\frac{H}{H_c} \quad (82)$$

$$Z_{ss}(\lambda_0) = \sqrt{2}\left(\frac{1}{2} + \frac{1}{8\ln(H_0/H)}\right) \quad (83)$$

To obtain the conformal dimensions in terms of small magnetic field we use Eqns. (80) to (83) on Eqns. (7) and (8). Note that,

$$(\det Z)^2 = (Z_{cc}^2 + Z_{cs}^2)(Z_{ss}^2 + Z_{sc}^2) - (Z_{cc}Z_{sc} + Z_{cs}Z_{ss})^2$$
$$\det Z = Z_{ss} \quad (84)$$

Therefore, the magnetic field dependence of the conformal dimensions are given by

$$2\Delta_c^\pm(\Delta N, D) = \left(Z_{cc}D_c + D_s\left(\frac{1}{2} - \frac{2}{\pi^2}\frac{H}{H_c}\right) \pm \frac{Z_{ss}\Delta N_c}{2 Z_{ss}}\right)^2 + 2N_c^\pm$$

$$= \left((D_c + \tfrac{1}{2}D_s) \pm \tfrac{1}{2}\Delta N_c - \frac{2D_s}{\pi^2}\frac{H}{H_c}\right)^2 + 2N_c^\pm \qquad (85)$$

$$2\Delta_s^\pm = \frac{1}{2}\left\{D_s \pm \left(\Delta N_s - \Delta N_c\left(\frac{1}{2} - \frac{2}{\pi^2}\frac{H}{H_c}\right)\right)\right\}^2 + \frac{1}{4\ln(H_0/H)} \qquad (86)$$

$$\times \left\{D_s^2 + \left(\Delta N_s - \Delta N_c\left(\frac{1}{2} - \frac{2}{\pi^2}\frac{H}{H_c}\right)\right)^2\right\} + 2N_s^\pm$$

According to the principles of CFT, the general expression for correlation function contains factors from holons and spinons, given by (Parola and Sorella, 1990)

$$G(t,x) \approx \sum_{D_c, D_s} \frac{a_k(D_c, D_s)\exp(-2iD_c k_{F,\uparrow}x)\exp(-2i(D_c + D_s)k_{F,\downarrow}x)}{(x - iv_c t)^{2\Delta_c^+}(x + iv_c t)^{2\Delta_c^-}(x - iv_s t)^{2\Delta_s^+}(x + iv_s t)^{2\Delta_s^-}} \qquad (87)$$

Where $k_{F,\uparrow}$ and $k_{F,\downarrow}$ are the Fermi momenta for electrons with spin up and down, respectively. $v_c$ and $v_s$ are the Fermi velocities of holon and spinon and $a_k$ are constant coefficients.

### 3. Correlation Functions in Magnetic Field

We now use the results obtained in the last section to obtain the magnetic field dependence of the unusual exponents of the electron field correlation function and density-density correlation function by setting the non-negative integer $N_{c,s}^\pm = 1$. First we consider the electron field correlation function with up-spin which originates from the quantum numbers $(D_c, D_s) = $ (1/2,-1/2), (3/2,-3/2), (5/2,-5/2), (5/2,-5/2), (7/2,-7/2), (9/2,-9/2), (11/2,-11/2), (13/2,-13/2), $\Delta N_c = 1$ and $\Delta N_s = 0$. Therefore, the corresponding conformal dimensions for $(D_c, D_s) = (1/2, -1/2)$ are

$$2\Delta_c^\pm = \left(\frac{1}{4} \pm \frac{1}{2} + \frac{1}{\pi^2}\frac{H}{H_c}\right)^2 + 2 \qquad (88)$$

$$2\Delta_c^+ = \frac{41}{16} - \frac{3}{2\pi^2}\frac{H}{H_c} + \frac{1}{\pi^4}\left(\frac{H}{H_c}\right)^2$$

$$2\Delta_c^- = \frac{33}{16} + \frac{1}{2\pi^2}\frac{H}{H_c} + \frac{1}{\pi^4}\left(\frac{H}{H_c}\right)^2 \qquad (89)$$

$$2\Delta_s^\pm = \frac{1}{2}\left\{-\frac{1}{2} \pm \left(-\frac{1}{2} + \frac{2}{\pi^2}\frac{H}{H_c}\right)\right\}^2 + \frac{1}{8\ln(H_0/H)}$$

$$- \frac{1}{2\pi^2}\frac{H}{H_c \ln(H_0/H)} + \frac{1}{\pi^4}\frac{H^2}{H_c^2 \ln(H_0/H)} + 2 \quad (90)$$

$$2\Delta_s^+ = \frac{5}{2} + \frac{2}{\pi^2}\frac{H}{H_c} - \frac{1}{16\ln(H_0/H)}$$

$$2\Delta_s^- = 2 - \frac{1}{16\ln(H_0/H)} \qquad (91)$$

Where the contributions from $(H/H_c)^2$ and terms of order $O(H/H_c \ln(H_0/H))$ are neglected. Using Eqns. (89) and (91) on (87), we obtain

$$\frac{a_1 \exp(-ik_{F,\uparrow}x)}{|x + iv_c t|^{\theta_{c1}}|x + iv_s t|^{\theta_{s1}}} \qquad (92)$$

The critical exponent is given by

$$\theta_{ci,si} = 2\Delta_{c,s}^+ + 2\Delta_{c,s}^- \qquad (93)$$

This implies that

$$\theta_{c1} = \frac{37}{8} - \frac{1}{\pi^2}\frac{H}{H_c} \qquad (94)$$

and

$$\theta_{s1} = \frac{9}{2} + \frac{2}{\pi^2}\frac{H}{H_c} - \frac{1}{8\ln(H_0/H)} \qquad (95)$$

Next, we obtain the conformal dimensions for $(D_c, D_s) = (3/2, -3/2)$ as

$$2\Delta_c^\pm = \left(\frac{3}{4} \pm \frac{1}{2} + \frac{3}{\pi^2}\frac{H}{H_c}\right)^2 + 2 \qquad (96)$$

$$2\Delta_c^+ = \frac{57}{16} + \frac{15}{2\pi^2}\frac{H}{H_c}$$

$$2\Delta_c^- = \frac{33}{16} + \frac{3}{2\pi^2}\frac{H}{H_c} \qquad (97)$$

$$2\Delta_s^\pm = \frac{1}{2}\left\{-\frac{3}{2} \pm \left(-\frac{1}{2} + \frac{2}{\pi^2}\frac{H}{H_c}\right)\right\}^2$$

$$+ \frac{1}{4\ln(H_0/H)}\left\{\frac{5}{2} - \frac{2}{\pi^2}\frac{H}{H_c} + \frac{4}{\pi^4}\left(\frac{H}{H_c}\right)^2\right\} + 2 \qquad (98)$$

$$2\Delta_s^+ = 4 - \frac{4}{\pi^2}\frac{H}{H_c} + \frac{5}{8\ln(H_0/H)} \qquad (99)$$

$$2\Delta_s^- = \frac{5}{2} + \frac{2}{\pi^2}\frac{H}{H_c} + \frac{5}{8\ln(H_0/H)}$$

Using Eqns. (97) and (99) on (87), we obtain

$$\frac{a_2 \exp(-3ik_{F,\uparrow}x)}{|x+iv_ct|^{\theta_{c2}}|x+iv_st|^{\theta_{s2}}} \qquad (100)$$

The critical exponent

$$\theta_{c2} = \frac{45}{8} + \frac{9}{\pi^2}\frac{H}{H_c}, \qquad (101)$$

and

$$\theta_{s2} = \frac{13}{2} - \frac{2}{\pi^2}\frac{H}{H_c} + \frac{5}{4\ln(H_0/H)} \qquad (102)$$

Next, for $(D_c, D_s) = (5/2, -5/2)$, we obtain the conformal dimensions as

$$2\Delta_c^\pm = \left(\frac{5}{4} \pm \frac{1}{2} + \frac{5}{\pi^2}\frac{H}{H_c}\right)^2 + 2 \qquad (103)$$

$$2\Delta_c^+ = \frac{81}{16} + \frac{35}{2\pi^2}\frac{H}{H_c}$$

$$2\Delta_c^- = \frac{41}{16} + \frac{15}{2\pi^2}\frac{H}{H_c} \qquad (104)$$

$$2\Delta_s^\pm = \frac{1}{2}\left\{-\frac{5}{2} \pm \left(-\frac{1}{2} + \frac{2}{\pi^2}\frac{H}{H_c}\right)\right\}^2$$

$$+ \frac{1}{4\ln(H_0/H)}\left\{\frac{13}{2} + \frac{2}{\pi^2}\frac{H}{H_c} - \frac{4}{\pi^4}\left(\frac{H}{H_c}\right)^2\right\} + 2 \qquad (105)$$

$$2\Delta_s^+ = \frac{13}{2} - \frac{6}{\pi^2}\frac{H}{H_c} + \frac{13}{8\ln(H_0/H)} \qquad (106)$$

$$2\Delta_s^- = 4 + \frac{4}{\pi^2}\frac{H}{H_c} + \frac{13}{8\ln(H_0/H)}$$

Using Eqns. (104) and (106) on (87), we obtain

$$\frac{a_3 \exp(-5ik_{F,\uparrow}x)}{|x+iv_ct|^{\theta_{c3}}|x+iv_st|^{\theta_{s3}}} \qquad (107)$$

The critical exponent

$$\theta_{c3} = \frac{61}{8} + \frac{25}{\pi^2}\frac{H}{H_c} \qquad (108)$$

and

$$\theta_{s3} = \frac{21}{2} - \frac{2}{\pi^2}\frac{H}{H_c} + \frac{13}{4\ln(H_0/H)} \qquad (109)$$

For $(D_c,\ D_s) = (7/2,\ -7/2)$, we obtain the conformal dimensions as

$$2\Delta_c^\pm = \left(\frac{7}{4} \pm \frac{1}{2} + \frac{7}{\pi^2}\frac{H}{H_c}\right)^2 + 2 \qquad (110)$$

$$2\Delta_c^+ = \frac{113}{16} + \frac{63}{2\pi^2}\frac{H}{H_c}$$

$$2\Delta_c^- = \frac{57}{16} + \frac{35}{2\pi^2}\frac{H}{H_c} \qquad (111)$$

$$2\Delta_s^+ = 10 - \frac{8}{\pi^2}\frac{H}{H_c} + \frac{25}{8\ln(H_0/H)}$$

$$2\Delta_s^- = \frac{13}{2} + \frac{6}{\pi^2}\frac{H}{H_c} + + \frac{25}{8\ln(H_0/H)} \qquad (112)$$

$$2\Delta_s^+ = 10 - \frac{8}{\pi^2}\frac{H}{H_c} + \frac{25}{8\ln(H_0/H)}$$

$$2\Delta_s^- = \frac{13}{2} + \frac{6}{\pi^2}\frac{H}{H_c} + + \frac{25}{8\ln(H_0/H)} \qquad (113)$$

Using Eqns. (111) and (113) on (87), we obtain

$$\frac{a_4 \exp(-7ik_{F,\uparrow}x)}{|x+iv_ct|^{\theta_{c4}}|x+iv_st|^{\theta_{s4}}} \qquad (114)$$

The critical exponent

$$\theta_{c4} = \frac{85}{8} + \frac{49}{\pi^2}\frac{H}{H_c} \qquad (115)$$

and

$$\theta_{s4} = \frac{33}{2} - \frac{2}{\pi^2}\frac{H}{H_c} + \frac{25}{4\ln(H_0/H)} \qquad (116)$$

For $(D_c,\ D_s) = (9/2,\ -9/2)$, we obtain the conformal dimensions as

$$2\Delta_c^\pm = \left(\frac{9}{4} \pm \frac{1}{2} + \frac{9}{\pi^2}\frac{H}{H_c}\right)^2 + 2 \qquad (117)$$

$$2\Delta_c^+ = \frac{153}{16} + \frac{99}{2\pi^2}\frac{H}{H_c}$$

$$2\Delta_c^- = \frac{81}{16} + \frac{63}{2\pi^2}\frac{H}{H_c} \qquad (118)$$

$$2\Delta_s^\pm = \frac{1}{2}\left\{-\frac{9}{2} \pm \left(-\frac{1}{2} + \frac{2}{\pi^2}\frac{H}{H_c}\right)\right\}^2 + \frac{41}{8\ln(H_0/H)} + 2 \qquad (119)$$

$$2\Delta_s^+ = \frac{29}{2} - \frac{10}{\pi^2}\frac{H}{H_c} + \frac{41}{8\ln(H_0/H)}$$

$$2\Delta_s^- = 10 + \frac{8}{\pi^2}\frac{H}{H_c} + \frac{41}{8\ln(H_0/H)} \qquad (120)$$

Using Eqns. (118) and (120) on (87), we obtain

$$\frac{a_5 \exp(-9ik_{F,\uparrow}x)}{|x+iv_ct|^{\theta_{c5}}|x+iv_st|^{\theta_{s5}}} \qquad (121)$$

The critical exponent

$$\theta_{c5} = \frac{117}{8} + \frac{81}{\pi^2}\frac{H}{H_c} \qquad (122)$$

and

$$\theta_{s5} = \frac{49}{2} - \frac{2}{\pi^2}\frac{H}{H_c} + \frac{41}{4\ln(H_0/H)} \qquad (123)$$

For $(D_c, D_s) = (11/2, -11/2)$, we obtain the conformal dimensions as

$$2\Delta_c^\pm = \left(\frac{11}{4} \pm \frac{1}{2} + \frac{11}{\pi^2}\frac{H}{H_c}\right)^2 + 2 \qquad (124)$$

$$2\Delta_c^+ = \frac{201}{16} + \frac{143}{2\pi^2}\frac{H}{H_c}$$

$$2\Delta_c^- = \frac{113}{16} + \frac{99}{2\pi^2}\frac{H}{H_c} \qquad (125)$$

$$2\Delta_s^\pm = \frac{1}{2}\left\{-\frac{11}{2} \pm \left(-\frac{1}{2} + \frac{2}{\pi^2}\frac{H}{H_c}\right)\right\}^2 + \frac{61}{8\ln(H_0/H)} + 2 \qquad (126)$$

$$2\Delta_s^+ = 20 - \frac{12}{\pi^2}\frac{H}{H_c} + \frac{61}{8\ln(H_0/H)}$$

$$2\Delta_s^- = \frac{29}{2} + \frac{10}{\pi^2}\frac{H}{H_c} + \frac{61}{8\ln(H_0/H)} \qquad (127)$$

Using Eqns. (125) and (127) on (87), we obtain

$$\frac{a_6 \exp(-11ik_{F,\uparrow}x)}{|x+iv_ct|^{\theta_{c6}}|x+iv_st|^{\theta_{s6}}} \qquad (128)$$

The critical exponent

$$\theta_{c6} = \frac{157}{8} + \frac{121}{\pi^2}\frac{H}{H_c} \qquad (129)$$

and

$$\theta_{s6} = \frac{69}{2} - \frac{2}{\pi^2}\frac{H}{H_c} + \frac{61}{4\ln(H_0/H)} \qquad (130)$$

Finally, for $(D_c, D_s) = (13/2, -13/2)$, we obtain the conformal dimensions as

$$2\Delta_c^\pm = \left(\frac{13}{4} \pm \frac{1}{2} + \frac{13}{\pi^2}\frac{H}{H_c}\right)^2 + 2 \qquad (131)$$

$$2\Delta_c^+ = \frac{257}{16} + \frac{195}{2\pi^2}\frac{H}{H_c}$$

$$2\Delta_c^- = \frac{153}{16} + \frac{143}{2\pi^2}\frac{H}{H_c} \qquad (132)$$

$$2\Delta_s^\pm = \frac{1}{2}\left\{-\frac{13}{2} \pm \left(-\frac{1}{2} + \frac{2}{\pi^2}\frac{H}{H_c}\right)\right\}^2 + \frac{85}{8\ln(H_0/H)} + 2 \qquad (133)$$

$$2\Delta_s^+ = 20 + \frac{12}{\pi^2}\frac{H}{H_c} + \frac{85}{8\ln(H_0/H)}$$

$$2\Delta_s^- = \frac{53}{2} - \frac{14}{\pi^2}\frac{H}{H_c} + \frac{85}{8\ln(H_0/H)} \qquad (134)$$

Using Eqns. (132) and (134) on (87), we obtain

$$\frac{a_7 \exp(-13ik_{F,\uparrow}x)}{|x+iv_ct|^{\theta_{c7}}|x+iv_st|^{\theta_{s7}}} \qquad (135)$$

The critical exponent

$$\theta_{c7} = \frac{205}{8} + \frac{169}{\pi^2}\frac{H}{H_c} \qquad (136)$$

and

$$\theta_{s7} = \frac{93}{2} - \frac{2}{\pi^2}\frac{H}{H_c} + \frac{85}{4\ln(H_0/H)} \qquad (137)$$

Combining Eqns. (92), (100), (107), (114), (121), (128) and (135), we obtain the long-distance asymptotic form of the electron field correlation function with up-spin as

$$G^{\uparrow}(x,t) \approx \frac{a_1 \exp(-ik_{F,\uparrow}x)}{|x+iv_ct|^{\theta_{c1}}|x+iv_st|^{\theta_{s1}}} + \frac{a_2 \exp(-3ik_{F,\uparrow}x)}{|x+iv_ct|^{\theta_{c2}}|x+iv_st|^{\theta_{s2}}} + \frac{a_3 \exp(-5ik_{F,\uparrow}x)}{|x+iv_ct|^{\theta_{c3}}|x+iv_st|^{\theta_{s3}}} + \frac{a_4 \exp(-7ik_{F,\uparrow}x)}{|x+iv_ct|^{\theta_{c4}}|x+iv_st|^{\theta_{s4}}}$$
$$+ \frac{a_5 \exp(-9ik_{F,\uparrow}x)}{|x+iv_ct|^{\theta_{c5}}|x+iv_st|^{\theta_{s5}}} + \frac{a_6 \exp(-11ik_{F,\uparrow}x)}{|x+iv_ct|^{\theta_{c6}}|x+iv_st|^{\theta_{s6}}} + \frac{a_7 \exp(-13ik_{F,\uparrow}x)}{|x+iv_ct|^{\theta_{c7}}|x+iv_st|^{\theta_{s7}}} \qquad (138)$$

Lastly, we consider the density-density correlation function which originates from the quantum numbers $(D_c, D_s) = (-1,1), (-2,2), (-3,3), (-4,4)$, $\Delta N_c = \Delta N_s = 0$ and $N_{c,s}^{\pm} = 1$. Here the corresponding conformal dimensions for $(D_c, D_s) = (-1,1)$ are

$$2\Delta_c^+ = \frac{9}{4} + \frac{2}{\pi^2}\frac{H}{H_c}$$
$$2\Delta_c^- = \frac{9}{4} + \frac{2}{\pi^2}\frac{H}{H_c} \qquad (139)$$

$$2\Delta_s^+ = \frac{5}{2} + \frac{1}{4\ln(H_0/H)}$$
$$2\Delta_s^- = \frac{5}{2} + \frac{1}{4\ln(H_0/H)} \qquad (140)$$

Again contributions from $(H/H_c)^2$ are neglected. Using Eqns. (139) and (140) on (87), we obtain

$$\frac{a_1 \exp(2ik_{F,\uparrow}x)}{|x+iv_ct|^{\theta_{c1}}|x+iv_st|^{\theta_{s1}}} \qquad (141)$$

The critical exponents are given by

$$\theta_{c1} = \frac{9}{2} + \frac{4}{\pi^2}\frac{H}{H_c} \qquad (142)$$

and

$$\theta_{s1} = 5 + \frac{1}{2\ln(H_0/H)} \qquad (143)$$

For $(D_c, D_s) = (-2,2)$ the conformal dimensions are

$$2\Delta_c^+ = 3 + \frac{8}{\pi^2}\frac{H}{H_c}$$
$$2\Delta_c^- = 3 + \frac{8}{\pi^2}\frac{H}{H_c} \qquad (144)$$

$$2\Delta_s^+ = 4 + \frac{1}{4\ln(H_0/H)}$$
$$2\Delta_s^- = 4 + \frac{1}{4\ln(H_0/H)} \qquad (145)$$

Using Eqns. (144) and (145) on (87), we obtain

$$\frac{a_2 \exp(4ik_{F,\uparrow}x)}{|x+iv_ct|^{\theta_{c2}}|x+iv_st|^{\theta_{s2}}} \qquad (146)$$

The critical exponents are given by

$$\theta_{c2} = 6 + \frac{16}{\pi^2}\frac{H}{H_c} \qquad (147)$$

and

$$\theta_{s2} = 8 + \frac{2}{\ln(H_0/H)} \qquad (148)$$

Next, for $(D_c, D_s) = (-3,3)$ the conformal dimensions are

$$2\Delta_c^+ = \frac{17}{4} + \frac{18}{\pi^2}\frac{H}{H_c}$$

$$2\Delta_c^- = \frac{17}{4} + \frac{18}{\pi^2}\frac{H}{H_c} \tag{149}$$

$$2\Delta_s^+ = \frac{13}{2} + \frac{9}{4\ln(H_0/H)}$$

$$2\Delta_s^- = \frac{13}{2} + \frac{9}{4\ln(H_0/H)} \tag{150}$$

Using Eqns. (149) and (150) on (87), we obtain

$$\frac{a_3 \exp(6ik_{F,\uparrow}x)}{|x+iv_c t|^{\theta_{c3}} |x+iv_s t|^{\theta_{s3}}}, \tag{151}$$

$$\theta_{c3} = \frac{17}{2} + \frac{36}{\pi^2}\frac{H}{H_c} \tag{152}$$

and

$$\theta_{s3} = 13 + \frac{9}{2\ln(H_0/H)} \tag{153}$$

Finally, for $(D_c, D_s) = (-4, 4)$ the conformal dimensions are

$$2\Delta_c^+ = 6 + \frac{32}{\pi^2}\frac{H}{H_c}$$

$$2\Delta_c^- = 6 + \frac{32}{\pi^2}\frac{H}{H_c} \tag{154}$$

$$2\Delta_s^+ = 10 + \frac{4}{\ln(H_0/H)}$$

$$2\Delta_s^- = 10 + \frac{4}{\ln(H_0/H)} \tag{155}$$

Using Eqns. (154) and (155) on (87), we obtain

$$\frac{a_4 \exp(8ik_{F,\uparrow}x)}{|x+iv_c t|^{\theta_{c4}} |x+iv_s t|^{\theta_{s4}}}, \tag{156}$$

$$\theta_{c4} = 12 + \frac{64}{\pi^2}\frac{H}{H_c} \tag{157}$$

and

$$\theta_{s4} = 20 + \frac{8}{\ln(H_0/H)} \tag{158}$$

Combining Eqns. (141), (146), (151) and (156), we obtain the long-distance asymptotic form of the density-density correlation function as

$$G(x,t) \approx \frac{a_1 \exp(2ik_{F,\uparrow}x)}{|x+iv_c t|^{\theta_{c1}} |x+iv_s t|^{\theta_{s1}}} + \frac{a_2 \exp(4ik_{F,\uparrow}x)}{|x+iv_c t|^{\theta_{c2}} |x+iv_s t|^{\theta_{s2}}} + \frac{a_3 \exp(6ik_{F,\uparrow}x)}{|x+iv_c t|^{\theta_{c3}} |x+iv_s t|^{\theta_{s3}}} + \frac{a_4 \exp(8ik_{F,\uparrow}x)}{|x+iv_c t|^{\theta_{c4}} |x+iv_s t|^{\theta_{s4}}} \tag{159}$$

## 4. Correlation Function in Momentum Space

The electron field correlation function Eqn. (138) has singularities at the Fermi points $k_{F,\uparrow}$, $3k_{F,\uparrow}$, $5k_{F,\uparrow}$, $7k_{F,\uparrow}$, $9k_{F,\uparrow}$, $11k_{F,\uparrow}$ and $13k_{F,\uparrow}$ respectively. Therefore, at $k \approx k_{F,\uparrow}$, the momentum distribution is given by

$$\tilde{G}^\uparrow(k \approx k_{F,\uparrow}) \approx [\text{sgn}(k-k_{F,\uparrow})]^{2s} |k-k_{F,\uparrow}|^\nu$$

$$\approx \text{sgn}(k-k_{F,\uparrow}) |k-k_{F,\uparrow}|^\nu \tag{160}$$

The critical exponent

$$\nu = \theta_{c,1} + \theta_{s,1} - 1 = \frac{65}{8} + \frac{1}{\pi^2}\frac{H}{H_c}, \tag{161}$$

and
$$2s = 2(\Delta_c^+ - \Delta_c^- + \Delta_s^+ - \Delta_s^-) = 1 \qquad (162)$$

Here we neglect logarithmic field dependence. Eqn. (160) represents the momentum distribution function around $k_F$ for the electron field correlator. It exhibits a typical power-law behaviour of the TL liquid, with critical exponent given by Eqn. (161). This unusual exponent grows monotonically with magnetic field as shown in Figure 1, and at zero magnetic field $\nu \to 8.125$.

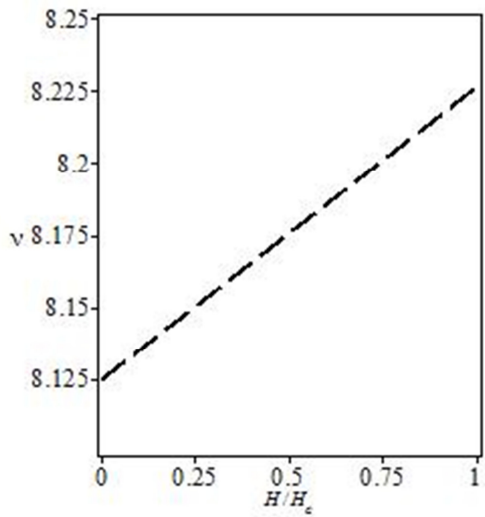

Figure 1. The critical exponent $\nu$ around the Fermi point $k_F$ as a function of magnetic field $H/H_c$ in the Hubbard model.

Another singularity is at $k \approx 3k_{F,\uparrow}$. The momentum distribution here is
$$\tilde{G}^\uparrow(k \approx 3k_{F,\uparrow}) \approx [\text{sgn}(k - 3k_{F,\uparrow})]^3 |k - 3k_{F,\uparrow}|^\nu, \qquad (163)$$

with critical exponent
$$\nu = \theta_{c,2} + \theta_{s,2} - 1 = \frac{89}{8} + \frac{7}{\pi^2}\frac{H}{H_c} \qquad (164)$$

Eqn. (163) shows typical power-law singularity of the TL liquid around the Fermi point $3k_F$, and the critical exponent Eqn. (164) grows monotonically with magnetic field as shown in Figure 2. At zero magnetic field goes to $\nu$ goes to $11.125$.

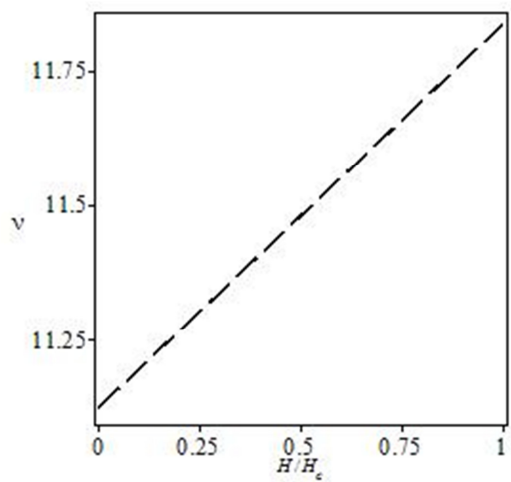

Figure 2. The critical exponent $\nu$ around the Fermi point $3k_F$ as a function of magnetic field $H/H_c$ in the Hubbard model

Next at $k \approx 5k_{F,\uparrow}$, the momentum distribution is given by

$$\tilde{G}^{\uparrow}(k \approx 5k_{F,\uparrow}) \approx [\text{sgn}(k - 5k_{F,\uparrow})]^5 \left|k - 5k_{F,\uparrow}\right|^\nu \quad (165)$$

with the unusual exponent

$$\nu = \theta_{c,3} + \theta_{s,3} - 1 = \frac{137}{8} + \frac{23}{\pi^2}\frac{H}{H_c} \quad (166)$$

Also, Eqn. (165) represents the momentum distribution function around the Fermi point $5k_F$ for the electron field correlator, and it exhibits a typical power-law behaviour of the TL liquid with critical exponent given by Eqn. (166). Here, at zero magnetic field the unusual exponent goes to 17.125 and it grows monotonically with growing magnetic field as shown in Figure 3.

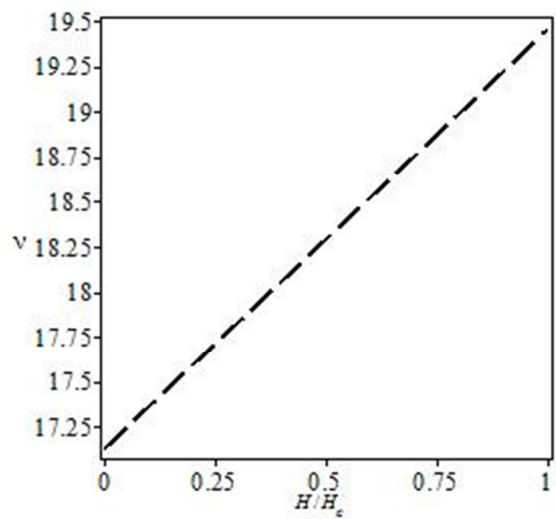

Figure 3. The critical exponent $\nu$ around the Fermi point $5k_F$ as a function of magnetic field $H/H_c$ in the Hubbard model

At $k \approx 7k_{F,\uparrow}$, the momentum distribution is

$$\tilde{G}^{\uparrow}(k \approx 7k_{F,\uparrow}) \approx [\text{sgn}(k - 7k_{F,\uparrow})]^7 \left|k - 7k_{F,\uparrow}\right|^\nu \quad (167)$$

and

$$\nu = \theta_{c,4} + \theta_{s,4} - 1 = \frac{209}{8} + \frac{47}{\pi^2}\frac{H}{H_c} \quad (168)$$

Eqn. (167) exhibits a typical power-law behaviour of the TL liquid around $7k_{F,\uparrow}$, with critical exponent given by Eqn. (168). This unusual exponent $\nu \to 26.125$ as the magnetic field goes to zero and increases monotonically with magnetic field as shown in Figure 4.

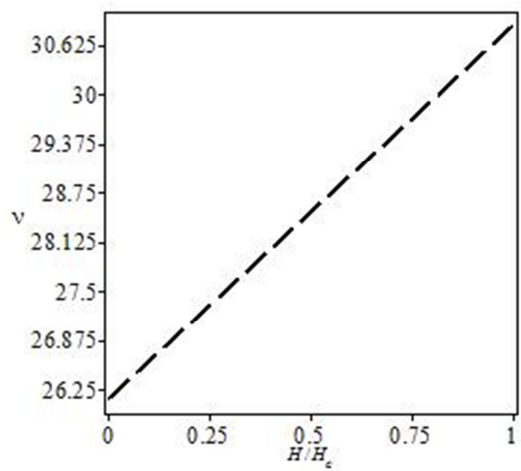

Figure 4. The critical exponent $\nu$ around the Fermi point $7k_F$ as a function of magnetic field $H/H_c$ in the Hubbard model

At $k \approx 9k_{F,\uparrow}$ the momentum distribution takes the form

$$\tilde{G}^\uparrow(k \approx 9k_{F,\uparrow}) \approx [\text{sgn}(k - 9k_{F,\uparrow})]^9 \left|k - 9k_{F,\uparrow}\right|^\nu, \quad (169)$$

with

$$\nu = \theta_{c,5} + \theta_{s,5} - 1 = \frac{305}{8} + \frac{79}{\pi^2}\frac{H}{H_c}, \quad (170)$$

and Eqn. (169) also exhibits a typical power-law behaviour of the TL liquid around the Fermi point $9k_{F,\uparrow}$, with critical exponent $\nu \to 38.125$ as the magnetic field goes to zero and increases monotonically with magnetic field as illustrated by Figure 5.

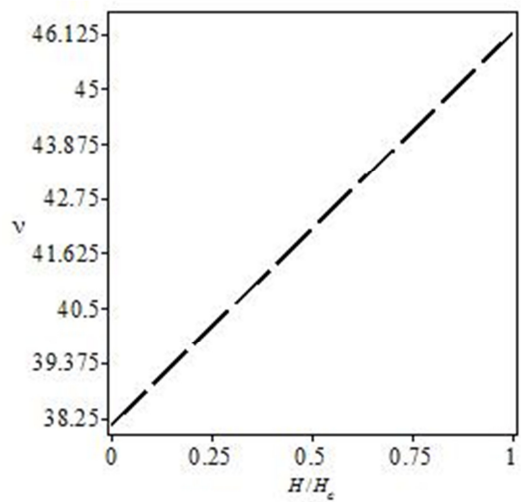

Figure 5. The critical exponent $\nu$ around the Fermi point $9k_F$ as a function of magnetic field $H/H_c$ in the Hubbard model

At $k \approx 11k_{F,\uparrow}$, the momentum distribution is

$$\tilde{G}^\uparrow(k \approx 11k_{F,\uparrow}) \approx [\text{sgn}(k - 11k_{F,\uparrow})]^{11} \left|k - 11k_{F,\uparrow}\right|^\nu \quad (171)$$

and

$$\nu = \theta_{c,6} + \theta_{s,6} - 1 = \frac{425}{8} + \frac{119}{\pi^2}\frac{H}{H_c} \quad (172)$$

Eqn. (171) exhibits a typical power-law behaviour of the TL liquid around $11k_{F,\uparrow}$, with critical exponent given by Eqn. (172). This unusual exponent $\nu \to 53.125$ as the magnetic field goes to zero and increases monotonically with magnetic field as shown in Figure 6.

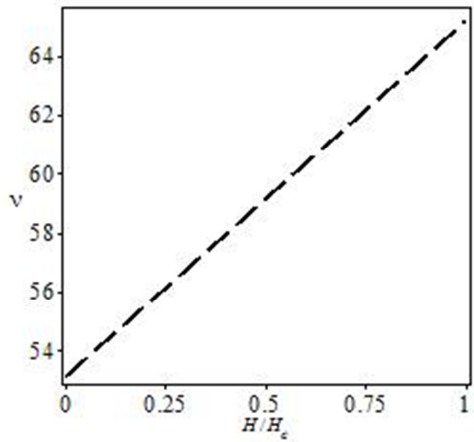

Figure 6. The critical exponent $v$ around the Fermi point $11k_F$ as a function of magnetic field $H/H_c$ in the Hubbard model

Finally, at $k \approx 13k_{F,\uparrow}$ the momentum distribution takes the form

$$\tilde{G}^{\uparrow}(k \approx 13k_{F,\uparrow}) \approx [\text{sgn}(k - 13k_{F,\uparrow})]^{13} |k - 13k_{F,\uparrow}|^{v}, \quad (173)$$

with

$$v = \theta_{c,7} + \theta_{s,7} - 1 = \frac{569}{8} + \frac{167}{\pi^2}\frac{H}{H_c}, \quad (174)$$

and Eqn. (173) also exhibits a typical power-law behaviour of the TL liquid around the Fermi point $13k_{F,\uparrow}$, with critical exponent $v \to 71.125$ as the magnetic field goes to zero and increases monotonically with magnetic field as depicted in Figure 7.

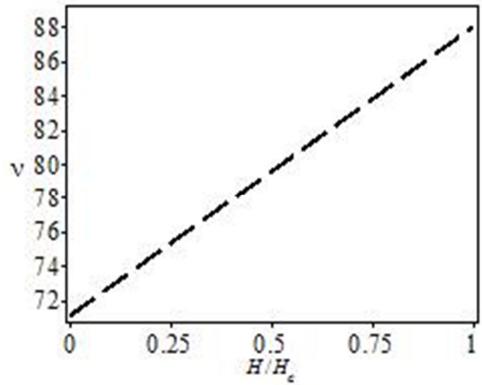

Figure 7. The critical exponent $v$ around the Fermi point $13k_F$ as a function of magnetic field $H/H_c$ in the Hubbard model

**5. Discussion**
In this study we have calculated the electron field and density-density correlation functions and their unusual exponents by using the non-negative integer $N^{\pm}_{c,s}$ characterizing particle-hole excitations as 1 in the 1D Hubbard model. Based on the principles of CFT we obtain expressions for the critical exponent that describes the long-distance behaviour of the correlation functions in coordinate and momentum space. The unusual behaviour of the exponent of correlation functions depends on the magnetic field and grows monotonically with increasing magnetic field. As the magnetic field goes to zero, the critical exponent goes to 8.125, 11.125, 17.125, 26.125 38.125, 53.125 and 71.125 around the Fermi points $k_F$, $3k_F$, $5k_F$, $7k_F$, $9k_F$, $11k_F$ and $13k_F$ respectively. It was observed that the $k_F$ part arises from the excitation of $(N^{\pm}_{c,s}, \Delta N_c, \Delta N_s, D_c, D_s) = (1,1,0,1/2,-1/2)$, the $3k_F$ part from $(N^{\pm}_{c,s}, \Delta N_c, \Delta N_s, D_c, D_s) = (1,1,0,3/2,-3/2)$, the $5k_F$ part from $(N^{\pm}_{c,s}, \Delta N_c, \Delta N_s, D_c, D_s) = (1,1,0,5/2,-5/2)$, the $7k_F$ part from $(N^{\pm}_{c,s}, \Delta N_c, \Delta N_s, D_c, D_s) = (1,1,0,7/2,-7/2)$, $9k_F$ part from $(N_{c,s}^{\pm}, \Delta N_c, \Delta N_s, D_c, D_s) = (1,1,0,9/2,-9/2)$, the $11k_F$ part from $(N_{c,s}^{\pm}, \Delta N_c, \Delta N_s, D_c, D_s) = (1,1,0,11/2,-11/2)$, and the $13k_F$ part from $(N_{c,s}^{\pm}, \Delta N_c, \Delta N_s, D_c, D_s) = (1,1,0,13/2,-13/2)$. This implies both holon and spinon excitations are responsible for the $k_F$, $3k_F$, $5k_F$, $7k_F$, $9k_F$, $11k_F$ and $13k_F$ oscillation parts respectively.

In conclusion, the electron field correlation function and the unusual exponents has been obtained around the Fermi points $k_F$, $3k_F$, $5k_F$, $7k_F$, $9k_F$, $11k_F$ and $13k_F$ respectively, and the density-density correlation function around $2k_F$, $4k_F$, $6k_F$ and $8k_F$. These results indicate that correlation functions of 1D Hubbard model exhibit power-law behaviour of Tomonaga-Luttinger liquid as the exponent $v$ changes monotonically with magnetic field. The properties around the Fermi points for $N_{c,s}^{\pm} = 1$ has led to greater understanding of the non-Fermi liquid TL unusual exponents of correlation functions in 1D correlated electron systems. This can also be investigated further to explore more physics around electron correlated systems.